\documentclass{article}

\usepackage{amsmath}
\usepackage{indentfirst}

\begin{document}

\newcommand{\unmedio}{{\scriptstyle\frac{1}{2}}}
\newcommand{\eff}{_{\text{eff}}}
\newcommand{\Infinity}{\infty}
\newcommand{\flip}{_{\text{flip}}}
\newcommand{\bos}{_{0\text{B}}}
\newcommand{\bosonic}{_{\text{bos}}}
\newcommand{\ferm}{_{0\text{F}}}
\newcommand{\trial}{_{\text{trial}}}
\newcommand{\true}{_{\text{true}}}

\newcommand{\sign}{\operatorname{sign}}
\newcommand{\Ci}{\operatorname{Ci}}
\newcommand{\tr}{\operatorname{tr}}

\newcommand{\PsiB}{\bar{\Psi}}   %Los campos
\newcommand{\phiH}{\hat{\phi}}
\newcommand{\etaH}{\hat{\eta}}
\newcommand{\chiB}{\bar{\chi}}
\newcommand{\xiH}{\hat{\xi}}
\newcommand{\zetaH}{\hat{\zeta}}
\newcommand{\vH}{\hat{v}}
\newcommand{\bH}{\hat{b}}

\newcommand{\AB}{\bar{A}_\mu}
\newcommand{\BB}{\bar{B}_\mu}
\newcommand{\AT}{\tilde{A}_\mu}
\newcommand{\BT}{\tilde{B}_\mu}

\newcommand{\slp}{\raise.15ex\hbox{$/$}\kern-.57em\hbox{$\partial$}}
\newcommand{\slA}{\raise.15ex\hbox{$/$}\kern-.63em\hbox{$A$}}

\newcommand{\difp}{\frac{d^2p}{(2\pi)^2}\,}

\newcommand{\bra}{\left\langle}
\newcommand{\ket}{\right\rangle}
\newcommand{\bracket}{\left\langle\,\right\rangle}

\newcommand{\D}{\mathcal{D}}
\newcommand{\N}{\mathcal{N}}
\newcommand{\Lag}{\mathcal{L}}
\newcommand{\V}{\mathcal{V}}
\newcommand{\Z}{\mathcal{Z}}

%\begin{frontmatter}

\thispagestyle{empty}
\title{Non local Thirring model with spin flipping interactions}
\author{An\'{\i}bal Iucci$^a$,  Kang Li$^{a,b}$ and Carlos M. Na\'on$^a$}
\date{}
\maketitle

\begin{abstract}

We extend a non local and non covariant version of the Thirring model in order to
describe a many-body system with spin-flipping interactions By introducing a model
with two fermion species we are able to avoid the use of non abelian bosonization
which is needed in a previous approach. We obtain a bosonized expression for the
partition function, describing the dynamics of the collective modes of this system. By
using the self-consistent harmonic approximation we found a formula for the gap of the
spin-charge excitations as functional of arbitrary electron-electron potentials.

\end{abstract}

\vspace{3cm} \noindent {\em Keywords} field theory, non local, functional
bosonization, many body, two dimensional, Luttinger liquid\\ \noindent {\em Pacs}:
11.10.Lm, 05.30.Fk\\ La Plata-Th 00/14

\noindent ---------------------------------------------------------------

\noindent $^a$ {\footnotesize Instituto de F\'{\i}sica La Plata, Departamento de
F\'{\i}sica, Facultad de Ciencias Exactas, Universidad Nacional de La Plata.  CC 67,
1900 La Plata, Argentina.}

\noindent $^b$ {\footnotesize Department of Physics, Zhejiang University, Hangzhou,
310027, P.R. of China.}

\noindent {\footnotesize emails: iucci@venus.fisica.unlp.edu.ar, kangli@mail.hz.zj.cn,
naon@venus.fisica.unlp.edu.ar.}

\newpage

\pagestyle{plain}

\section{Introduction}

In recent years there has been a surge of activity in the study of low-dimensional
field theories. In particular, research on the one-dimensional (1d) fermionic gas has
attracted a lot of attention. An interesting aspect of this system is the possibility
of having a deviation from the usual Fermi-liquid behavior. This phenomenon is known
as Luttinger-liquid behavior, characterized by spin-charge separation and by non
universal (interaction dependent) power-law correlation functions  \cite{Voit}. The
simplest theoretical framework that presents this feature is the Tomonaga-Luttinger
(TLM) model \cite{TLM}, a many-body system of right and left-moving particles
interacting through their charge densities. In recent papers \cite{NLT1} \cite{NLT2}
\cite{NLT3} an alternative, field theoretical approach was developed to consider this
problem. In these works a non local and non covariant version of the Thirring model
was introduced, in which the fermionic densities and currents are coupled through
bilocal, distance-dependent potentials. The resulting non local Thirring model (NLTM)
contains the TLM as a particular case. Although it provides an elegant framework to
analyze the 1d many-body problem, as it stands it is not very practical to consider
magnetic properties of Luttinger liquids. This is so because in the original
formulation of NLTM \cite{NLT1}, the description of spin-flipping (SF) processes,
based on non abelian bosonization, is quite involved. In this paper we show how to
circumvent this problem by introducing a Thirring-like model with two fermion species.
To this end we generalize the two-fermion model of Ref. \cite{Zinn-Justin}, originally
built as a local and covariant theory, to the case in which the interactions between
fermionic currents are mediated by bilocal functions. From the condensed matter
standpoint, our work can also be viewed as an extension of the work of Grinstein,
Minnhagen and Rosengren \cite{GMR} where a simplified version of the SF problem
(contained in our model) was first considered. Our procedure allows to get a bosonized
vacuum to vacuum functional by carefully extending previous formulations of massive
\cite{Coleman} \cite{Naon} and massive-like \cite{Li-Naon} models. The paper is
organized as follows. In Section 2 we introduce the above mentioned non local
two-fermion model and explain its relationship to previous non Abelian description. In
Section 3 we establish an equivalence between the initial fermionic partition function
and the one corresponding to a two-boson non local extension of the sine-Gordon model.
In Section 4 we employ a non local version of the self-consistent harmonic
approximation (SCHA) \cite{SCHA} in order to analyze the spectrum of the spin-density
modes. In particular we give, within this approximation, a closed formula for the gap
as functional of the non-contact forward-scattering (no SF) potentials. Finally, in
Section 5, we summarize the main points of our investigation and gather our
conclusions. In the Appendix we show how to compute the gap of the spin-charge sector
by means of the harmonic approximation.

\section{The model and its relationship to a previous non Abelian description}

 In this Section we introduce a non local version of the Thirring model which incorporates
electronic spin by considering two fermion species, where each species represents a
spin state. Our initial (Euclidean) action is

\begin{multline}\label{eq:model}
S = \int d^2x\,\PsiB^a i \slp \Psi^a - \frac{g^2}{2}\int d^2x\,d^2y\, j_\mu^a(x)
V^{ab}_{(\mu)}(x-y)j_\mu^b(y)\\ - g_s\int d^2x\,d^2y\,\PsiB^1 (x)\gamma_\mu\Psi^2
(x)U_{(\mu)}(x-y)\PsiB^2(y)\gamma_\mu\Psi^1(y)
\end{multline}
where $a,b=1,2$, with $1=\uparrow ,2=\downarrow$, the currents $j^a_\mu$ are the usual
fermion currents

\begin{equation}
j_\mu^1=\PsiB^1\gamma_\mu\Psi^1,\;\;\;\;
j_\mu^2=\PsiB^2\gamma_\mu\Psi^2,
\end{equation}

\noindent and the matrices $V_{(\mu)}^{ab}$ have the form

\begin{equation}
V_{(\mu)} = \frac{1}{2}
\begin{pmatrix}
      V^c_{(\mu)}+V^s_{(\mu)} & V^c_{(\mu)}-V^s_{(\mu)} \\
      V^c_{(\mu)}-V^s_{(\mu)} & V^c_{(\mu)}+V^s_{(\mu)} \\
\end{pmatrix},
\end{equation}

\noindent where $V^c_{(\mu)}$ and $V^s_{(\mu)}$ are coupling potentials related to
Solyom's \cite{Solyom} $g_2$ and $g_4$ forward scattering couplings. These functions
are related to current-current interactions which do not flip the spins. The functions
$U_{(\mu)}(x-y)$ are the couplings governing SF processes. We have kept the constants
$g$ and $g_S$ in order to facilitate comparison with local versions of this model. For
instance, the case $g_S=0$ and $V^c_({\mu})=V^s_({\mu})=\delta^2(x-y)$ corresponds to
two decoupled, usual Thirring models.  The action (\ref{eq:model}) has manifest $U(1)$
chiral symmetry, i.e., it is invariant under the transformation $\Psi^a\rightarrow
e^{i\gamma_5 \theta}\Psi^a$, $\PsiB^a\rightarrow \PsiB^a e^{i\gamma_5 \theta}$.
Fermion spin conservation is also preserved in this theory.

The theory defined above is similar to the one described by ZJ \cite{Zinn-Justin}.
There are, however, two important differences with that model. First of all, our model
takes into account the possible long-range nature of the potentials, whereas ZJ's
model is local. On the other hand ZJ's model does not include $g_4$-like terms
associated to scattering processes involving just one fermionic (left or right)
branch, neither for SF nor for ordinary diagrams.

Concerning the relationship of our action to previous condensed-matter inspired
models, we should mention the pioneering works of Luther and Emery
\cite{Luther-Emery}, and Grinstein, Minnhagen and Rosengren \cite{GMR}. The first
authors introduced the so called backscattering model. Although this system has in
principle no SF, in the local limit their backscattering diagrams coincide with the
ones of spin-changing processes. Grinstein et al. included from the beginning SF
interactions taking into account a Coulombian potential. Thus, their model is non
local, but they took the same potential for all kinds of diagrams (SF and ordinary).
Besides, in order to establish the equivalence between their theory and a Coulomb gas
system the authors considered again a local limit. These authors did not include
$g_4$-like terms either.

Let us now show that the action (\ref{eq:model}) can be written in an alternative way.
Consider the U(N) currents

\begin{equation}
J^\alpha_\mu=\PsiB\gamma_\mu\lambda^\alpha\Psi
\end{equation}

\noindent with

\begin{align}
\lambda^0&=\unmedio I \\ \lambda^j&=t^j,
\end{align}

\noindent $t^j$ being the SU(N) generators normalized according to

\begin{equation}
\tr{t^it^j}=\unmedio\delta^{ij}.
\end{equation}

With these currents, one can define a non local, chiral invariant, U(N) Gross-Neveu
model, with action given by

\begin{multline}\label{eq:grossneveu}
S=\int d^2x\, \PsiB i\slp \Psi - \int
d^2x\,d^2y\,J^\alpha_\mu(x)\V^{\alpha\beta}_{(\mu)}(x-y)J^\beta_\mu(y),\\
\alpha,\beta=0,1,...,N^2-1
\end{multline}

\noindent where $\V_{(\mu)}$ are $N^2\times N^2$ symmetric matrices that weight the
interaction. Taking $N=2$, the actions (\ref{eq:model}) and (\ref{eq:grossneveu}) are
equal provided that the matrices $\V_{(\mu)}$ can be written as

\begin{equation}
\V_{(\mu)} =
\begin{pmatrix}
      g^2 V^c_{(\mu)} & 0 & 0 & 0 \\
      0 & g_s U_{(\mu)} & 0 & 0  \\
      0 & 0 & g_s U_{(\mu)} & 0  \\
      0 & 0 & 0 & g^2 V^s_{(\mu)} \\
\end{pmatrix}.
\end{equation}

This non Abelian model was considered in Ref. \cite{NLT1}. The bosonized effective
action obtained by non Abelian bosonization led to a WZW functional, which was not
easy to deal with in order to analyze the physical spectrum. In the next Sections we
will show how this task is greatly simplified by starting from (\ref{eq:model})
instead of (\ref{eq:grossneveu}) and combining path-integral {\em Abelian}
bosonization and the self consistent harmonic approximation.

%%%%%%%%%%%%%%%%%%%%%%%%%%%%%%%%%%%%%%%%%%%%%%%%%%%%%%%%%%%%%%%%%%%%%%%%%%%%%%%%%%%%%%%%%%%%%%%%%%%%%%%%%%%%%%%%%%%%%%%%%%%
%%%%%%%%%%%%%%%%%%%%%%%%%%%%%%%%%%%%%%%%%%%%%%%%%%%%%%%%%%%%%%%%%%%%%%%%%%%%%%%%%%%%%%%%%%%%%%%%%%%%%%%%%%%%%%%%%%%%%%%%%%%
\section{The equivalent bosonic action}

We start by considering the partition function

\begin{equation}
\Z=\N\int\D\PsiB^a\D\Psi^a\,e^{-S},
\end{equation}

\noindent where $\N$ is a normalization constant. It is convenient to write

\begin{equation}
S=S_0 + S\flip,
\end{equation}

\noindent where

\begin{equation}
S_0=\int d^2x\,\PsiB^a i \slp \Psi^a - \frac{g^2}{2}\int d^2x\,d^2y\, j_\mu^a(x) V^{ab}_{(\mu)}(x-y)j_\mu^b(y)
\end{equation}

\noindent and

\begin{equation}\label{eq:flip}
S\flip=-g_s\int d^2x\,d^2y\,\PsiB^1(x)\gamma_\mu\Psi^2(x)
U_{(\mu)}(x-y)\PsiB^2(y)\gamma_\mu\Psi^1(y).
\end{equation}

\noindent The reason for this separation lies on the fact that $S_0$ contains all the
interaction terms that have separate chiral invariance for each species of fermions
(spin states). They are Thirring-like interactions and will be treated in much the
same way as the massless Thirring action, i.e. they will be transformed into free, non
local action terms. Concerning the second term, it has no separate chiral invariance,
and will be expanded in perturbative series like the mass term in the Thirring model.

It is easy to show that the introduction of auxiliary vector fields $A_\mu^a$ allows
to write

\begin{equation}
\Z=\N'\int\D\PsiB^a\D\Psi^a\D A_\mu^a\exp\left[-\int
d^2x\,\PsiB^a\left(i\slp + g\slA^a\right)\Psi^a - S[A] -
S\flip\right],
\end{equation}

\noindent where $\N'$ is a new constant (See \cite{NLT1} for details) and

\begin{equation}
S[A]=\frac{1}{2}\int d^2x\,d^2y\,\left(V^{-1}_{(\mu)}\right)^{ab}(x-y)A_\mu^a(x)A_\mu^b(y),
\end{equation}

\noindent with $\left(V^{-1}_{(\mu)}\right)^{ab}$ defined through the following equation:

\begin{equation}\label{eq:bV}
\int d^2y\,\left(V^{-1}_{(\mu)}\right)^{ab}(x-y)V_{(\mu)}^{bc}(y-z)=\delta^{(2)}(x-z)\delta^{ac}.
\end{equation}

\noindent We now decompose $A_\mu^a$ in longitudinal and transverse pieces

\begin{equation}
A_\mu^a(x)=\epsilon_{\mu\nu}\partial_{\nu}\phi^a(x)+ \partial_\mu\eta^a(x),
\end{equation}

\noindent where $\phi^a$ and $\eta^a$ are scalar fields. We also perform a change in
the fermionic fields

\begin{equation} \label{f1}
\Psi^a(x) = e^{-g[\gamma_5 \phi^a(x) - i \eta^a(x)]}\chi^a(x)
\end{equation}

\begin{equation} \label{f2}
\PsiB^a(x) = \chiB^a(x) e^{-g[\gamma_5 \phi^a(x) + i\eta^a(x)]}
\end{equation}

\noindent whose Jacobian, as it is well-known, yields a kinetic term for the $\phi^a$
fields. One thus gets

\begin{equation}
\Z=\N\int\D\chiB^a\D\chi^a\D\phi^a\D\eta^a e^{-S\eff},
\end{equation}

\noindent being $S\eff$ a sum of three pieces:

\begin{equation}
S\eff=S\ferm+S\bos+S\flip
\end{equation}

\noindent where

\begin{equation}
S\ferm=\int d^2x\,\left(\chiB^1 i\slp \chi^1 + \chiB^2 i\slp \chi^2\right),
\end{equation}

\begin{multline}
S\bos=\frac{g^2}{2\pi}\int d^2x\,\left[(\partial_\mu\phi^1)^2+(\partial_\mu\phi^2)^2\right]
+\frac{1}{2}\int d^2x\,d^2y\,\left(V^{-1}_{(\mu)}\right)^{ab}(x-y)\\
\times\left[\epsilon_{\mu\nu}\epsilon_{\mu\rho}\partial_\nu\phi^a(x)\partial_\rho\phi^b(y)
+\partial_\mu\eta^a(x)\partial_\mu\eta^b(y) +
2\epsilon_{\mu\nu}\partial_\nu\phi^a(x)\partial_\mu\eta^b(y)\right],
\end{multline}

\noindent and $S\flip$ is the same SF interaction term already defined in equation
(\ref{eq:flip}). Concerning this last term, from now on we shall restrict our study to
the case of contact SF interactions:

\begin{equation}
U(x-y)_{(0)}=U(x-y)_{(1)}=\delta^{(2)}(x-y),
\end{equation}

\noindent and by performing a Fierz transformation followed by the chiral change defined in equations (\ref{f1}) and
(\ref{f2}), we can write it in the form

\begin{multline}
S\flip=2 g_s\int d^2x\Big[
e^{-2g(\phi^1-\phi^2)}\chiB^1\frac{1+\gamma_5}{2}\chi^1\cdot
\chiB^2\frac{1-\gamma_5}{2}\chi^2 \\+
e^{2g(\phi^1-\phi^2)}\chiB^1\frac{1-\gamma_5}{2}\chi^1\cdot
\chiB^2\frac{1+\gamma_5}{2}\chi^2\Big].
\end{multline}

Now we are ready to make an expansion of the partition function taking $g_s$ as
perturbative parameter:

\begin{multline}
\Z=\N\int\D\phi^a\D\eta^a e^{-S\bos}
\sum_{n=0}^\Infinity\frac{(-2g_s)^n}{n!}\int\left(\prod_{i=1}^n d^2x_i\right)\\
\times\bigg\langle\prod_{i=1}^n\Big[e^{-2g(\phi^1-\phi^2)}\chiB^1\frac{1+\gamma_5}{2}\chi^1\cdot
\chiB^2\frac{1-\gamma_5}{2}\chi^2 \\+
e^{2g(\phi^1-\phi^2)}\chiB^1\frac{1-\gamma_5}{2}\chi^1\cdot
\chiB^2\frac{1+\gamma_5}{2}\chi^2\Big]\bigg\rangle\ferm
\end{multline}

\noindent where $\bracket\ferm$ means v.e.v. in a theory with action $S\ferm$. Only
mean values involving an equal number of factors of the form
$\unmedio\chiB^a(1+\gamma_5)\chi^a$ and $\unmedio\chiB^a(1-\gamma_5)\chi^a$ (no sum
over repeated indices is implied in these expressions) are non zero, and thus the
partition function can be written in the form

\begin{multline}
\Z=\N\int\D\phi^a\D\eta^a\, e^{-S\bos}\\
\times\sum_{n=0}^\Infinity\frac{(2g_s)^{2n}}{(n!)^2}\int\left(\prod_{i=1}^n
d^2x_i\,d^2y_i\,
e^{-2g\left[\phi^1(x_i)-\phi^2(x_i)-\phi^1(y_i)+\phi^2(y_i)\right]}\right)
\\ \times \bra\prod_{i=1}^n
\chiB^1(x_i)\frac{1+\gamma_5}{2}\chi^1(x_i)
\chiB^1(y_i)\frac{1-\gamma_5}{2}\chi^1(y_i)\ket\ferm\\
\times\bra\prod_{i=1}^n
\chiB^2(x_i)\frac{1-\gamma_5}{2}\chi^2(x_i)
\chiB^2(y_i)\frac{1+\gamma_5}{2}\chi^2(y_i)\ket\ferm.
\end{multline}

\noindent The next step is to introduce two local massless scalar fields $\vartheta^a$
to be associated with the free fermions  $\chiB^a$ and $\chi^a$. This trick allows to
replace the fermionic mean values in the above equation by their ultraviolet
regularized bosonic counterparts, which yields

\begin{multline}
\Z=\N\int\D\phi^a\D\eta^a\,
e^{-S\bos}\\\times\sum_{n=0}^\Infinity\frac{(2g_s)^{2n}}{(n!)^2}\int\left(\prod_{i=1}^n
d^2x_i\,d^2y_i
e^{-2g\left[\phi^1(x_i)-\phi^2(x_i)-\phi^1(y_i)+\phi^2(y_i)\right]}\right)
\\\times \left(\frac{i\Lambda}{2\pi}\right)^{4n} \bra\prod_{i=1}^n
e^{i\sqrt{4\pi}\left[\vartheta^1(x_i)-\vartheta^1(y_i)\right]}\ket_{0,\vartheta^1}
 \bra\prod_{i=1}^n
e^{i\sqrt{4\pi}\left[\vartheta^2(x_i)-\vartheta^2(y_i)\right]}\ket_{0,\vartheta^2}.
\end{multline}

We now use the fact that the infrared divergences of the $\vartheta^a$ field's
propagator provide a neutrality condition for v.e.v's of vertex operators. This allows
to rearrange the perturbative series in a non-trivial way. One is then led to the
completely bosonized action $S_{\text{bos}}$

\begin{equation}
S\bosonic=S\bos + \int d^2x\,
\left\{\frac{1}{2}(\partial_\mu\vartheta^a)^2
-\frac{g_s\Lambda^2}{\pi^2}
\cos\left[2ig(\phi^1-\phi^2)+\sqrt{4\pi}(\vartheta^1+\vartheta^2)\right]\right\}.
\end{equation}

At this point it is interesting to observe that there is a set of changes of variables that allows to express the action
in a very suggestive way. Indeed, writing

\begin{equation}
\theta=\frac{1}{\sqrt{2}}(\vartheta^1+\vartheta^2)
\end{equation}

\begin{equation}
\tilde{\theta}=\frac{1}{\sqrt{2}}(\vartheta^1-\vartheta^2)
\end{equation}

\begin{equation}\label{scSep1}
\phi^{1,2}=\frac{1}{\sqrt{2}}(\phi_c\pm\phi_s)
\end{equation}

\begin{equation}\label{scSep2}
\eta^{1,2}=\frac{1}{\sqrt{2}}(\eta_c\pm\eta_s)
\end{equation}

\noindent where the plus(minus) sign corresponds to the pair
${\phi^1,\eta^1}$(${\phi^2,\eta^2}$), one sees that the field $\tilde{\theta}$
completely decouples from the others and can then be integrated, and the
bosonized action becomes
\begin{equation}
S_{\text{bos}}=S_c+S_s,
\end{equation}
where

\begin{multline}
S_c=\frac{g^2}{2\pi}\int d^2x\,(\partial_\mu\phi_c)^2
+\frac{1}{2}\int d^2x\,d^2y\,\left(V^c_{(\mu)}\right)^{-1}(x-y)\\
\times\left[\epsilon_{\mu\nu}\epsilon_{\mu\rho}\partial_\nu\phi_c(x)\partial_\rho\phi_c(y)
+\partial_\mu\eta_c(x)\partial_\mu\eta_c(y) +
2\epsilon_{\mu\nu}\partial_\nu\phi_c(x)\partial_\mu\eta_c(y)\right],
\end{multline}

\noindent and

\begin{multline}\label{eq:SpinAction}
S_s=\frac{g^2}{2\pi}\int d^2x\,(\partial_\mu\phi_s)^2
+\frac{1}{2}\int d^2x\,d^2y\,\left(V^s_{(\mu)}\right)^{-1}(x-y)\\
\times\left[\epsilon_{\mu\nu}\epsilon_{\mu\rho}\partial_\nu\phi_s(x)\partial_\rho\phi_s(y)
+\partial_\mu\eta_s(x)\partial_\mu\eta_s(y) +
2\epsilon_{\mu\nu}\partial_\nu\phi_s(x)\partial_\mu\eta_s(y)\right]\\
+\int d^2x\left[\frac{1}{2}(\partial_\mu\theta)^2
-\frac{g_s\Lambda^2}{\pi^2}
\cos\left(\sqrt{8}ig\phi_s+\sqrt{8\pi}\theta\right)\right]
\end{multline}

\noindent with the functions $\left(V^{c,s}_{(\mu)}\right)^{-1}$ defined as

\begin{equation}
\int d^2y\,\left(V^{c,s}_{(\mu)}\right)^{-1}(x-y)
V^{c,s}_{(\mu)}(y-z) = \delta^{(2)}(x-z).
\end{equation}

This, in turn, leads to the factorization of the partition function in the form
$\Z=\Z_c \Z_s$. This result is a clear manifestation of the so called spin-charge
separation, a typical feature of Luttinger liquids \cite{Luther-Emery} \cite{GMR}.
$\Z_c$ is the partition function associated to charge density excitations. It
coincides with the spinless NLTM studied in reference \cite{NLT1}. $\Z_s$ describes
spin-density excitations. The action $S_s$ corresponds to a certain non local
sine-Gordon model similar to the one previously considered in \cite{Li-Naon}. In the
next Section we shall derive an expression for the gap of its spectrum.

Since our main goal is to analyze the spin-charge sector, from now on we shall focus
our attention to $S_s$. Going to momentum space (with the exception of the cosine
term, whose transformed expression is not very illuminating) one has

\begin{multline}
S_s=\int\difp\left[\phiH(p)\phiH(-p)A(p)+\etaH(p)\etaH(-p)B(p) +
\phiH(p)\etaH(-p)C(p)\right]\\+\int\difp\frac{p^2}{2}\theta(p)\theta(-p)
-\frac{g_s\Lambda^2}{\pi^2}\int d^2x\,
\cos\left(\sqrt{8}ig\phi_s+\sqrt{8\pi}\theta\right)
\end{multline}
with

\begin{equation}\label{eq:A}
A(p)=\frac{g^2}{2\pi} +\frac{1}{2}\left[\left(V^s_{(0)}\right)^{-1}(p)p_1^2
+\left(V^s_{(1)}\right)^{-1}(p)p_0^2\right],
\end{equation}

\begin{equation}\label{eq:B}
B(p)=\frac{1}{2}\left[\left(V^s_{(1)}\right)^{-1}(p)p_1^2
+\left(V^s_{(0)}\right)^{-1}(p)p_0^2\right],
\end{equation}

\begin{equation}\label{eq:C}
C(p)=\left[\left(V^s_{(0)}\right)^{-1}(p)-\left(V^s_{(1)}\right)^{-1}(p)\right]p_0p_1.
\end{equation}

In the above expressions $\phiH(p)$, $\etaH(p)$ and $\theta(p)$ are the Fourier
transforms of $\phi_s(x)$, $\eta_s(x)$ and $\theta(x)$ respectively.

%%%%%%%%%%%%%%%%%%%%%%%%%%%%%%%%%%%%%%%%%%%%%%%%%%%%%%%%%%%%%%%%%%%%%%%%%%%%%%%%%%%%%%%%%%%%%%%%%%%%%%%%%%%%%%%%%%%%%%%
%%%%%%%%%%%%%%%%%%%%%%%%%%%%%%%%%%%%%%%%%%%%%%%%%%%%%%%%%%%%%%%%%%%%%%%%%%%%%%%%%%%%%%%%%%%%%%%%%%%%%%%%%%%%%%%%%%%%%%%
\section{The spectrum of the spin-charge sector}

In order to obtain the approximate spectrum for the spin-charge sector of our model,
we will introduce a non local version of the self-consistent harmonic approximation
\cite{SCHA}. Basically, this amounts to replacing the so called {\em true} action
(\ref{eq:SpinAction}) by a {\em trial} action in which the cosine term is approximated
as
\begin{equation}
-\frac{g_s\Lambda^2}{\pi^2}
\cos\left(\sqrt{8}ig\phi_s+\sqrt{8\pi}\theta\right)\rightarrow \frac{\Omega^2}{2}
\left(\sqrt{8}ig\phi_s+\sqrt{8\pi}\theta\right)^2
\end{equation}
where the parameter $\Omega$ of the trial action can be variationally determined (See
the Appendix for details). Once this is done, it is straightforward to obtain the
spectrum. Indeed, going to momentum space, and back to real frequencies,
$p_0=i\omega$, $p_1=k$, the following equation is obtained:

\begin{equation}\label{eq:spectrum}
4\pi\Omega^2 + k^2(1+\frac{g^2}{\pi}V^s_{(0)})
-\omega^2(1+\frac{g^2}{\pi}V^s_{(1)})=0.
\end{equation}

Our result for $\Omega$ is

\begin{equation}\label{gap}
\Omega^2=\frac{g_s\Lambda^2}{\pi^2} e^{-4\pi I_0(\Omega)}
\end{equation}

where

\begin{equation}
I_{0}(\Omega)=\int\frac{d^2p}{(2\pi)^2}\frac{1}{p^2+\frac{g^2}{\pi}\left(p_0^2
V^s_{(1)}+p_1^2 V^s_{(0)}\right) + 8\pi\Omega^2}
\end{equation}

Equation (\ref{gap}) is one of our main results. It gives, within the self-consistent
harmonic approximation, a closed expression for the gap as functional of the
potentials $V^s_{(\mu)}(p)$.

At this stage it is interesting to consider the contact potential given by

\begin{equation}
V^s_{(0)}(p)=V^s_{(1)}(p)=1
\end{equation}

 In this case $I_0(\Omega)$ is infinite (See Appendix). After a suitable regularization one
obtains

\begin{equation}
\frac{\Omega^2}{\Lambda^2}=\frac{g_s}{\pi^2}
\left[1+\frac{(1+g^2/\pi)\Lambda^2}{8\pi\Omega^2}\right]^{\frac{-1}{1+g^2/\pi}}.
\end{equation}

\noindent This equation can be easily solved for $\Lambda\gg\Omega$ and
$\Omega\gg\Lambda$. The first case corresponds to $g_s\ll 1$, and the result is

\begin{equation}
\Omega^2=\frac{g_s\Lambda^2}{\pi^2}
\left[\frac{8g_s}{\pi(1+g^2/\pi)}\right]^{\pi/g^2}.
\end{equation}
Note that for the value $g^2=\pi$ the gap posses a linear dependence on the coupling.
It is interesting to observe that for this value, the action $S_s$ (see Appendix)
corresponds to a free massive fermion, with mass proportional to $g_s$. This is analog
to the behavior found on the Luther-Emery line \cite{Luther-Emery}. Despite this
similarity, the position of the line in our model differs slightly from theirs, but
since the models are not exactly equal such a disagreement is not unexpected. However,
we feel that the precise connection between both systems deserves a closer analysis
that will be undertaken in a future work.

The second case corresponds to $g_s\gg 1$, and we obtain

\begin{equation}
\Omega^2=\frac{g_s\Lambda^2}{\pi^2},
\end{equation}
which is independent of $g$. This is expected because in this limit the contribution
of the ordinary (without SF) terms are completely negligible.

Let us stress that our gap equation (\ref{gap}) provides a new tool to check the
validity of different potentials \cite{Iucci}. This is interesting because in
condensed-matter applications the assumption of short-range electron-electron
interactions works well for conductors in which the screening between adjacent chains
reduces the range of interactions within one chain \cite{Schulz1}. On the other hand,
as the dimensionality of the system decreases charge screening effects are expected to
become less important and the long-range interaction between electrons seems to play a
central role in determining the properties of the system. This assertion seems to be
confirmed by experiments in GaAs quantum wires \cite{quantum wires} and quasi-1d
conductors \cite{quasi-exp}. We hope to report on numerical solutions of the gap
equations for long-range potentials in a future contribution.

%%%%%%%%%%%%%%%%%%%%%%%%%%%%%%%%%%%%%%%%%%%%%%%%%%%%%%%%%%%%%%%%%%%%%%%%%%%%%%%%%%%%%%%%%%%%%%%%%%%%%%%%%%%%%%%%%%%%%%
%%%%%%%%%%%%%%%%%%%%%%%%%%%%%%%%%%%%%%%%%%%%%%%%%%%%%%%%%%%%%%%%%%%%%%%%%%%%%%%%%%%%%%%%%%%%%%%%%%%%%%%%%%%%%%%%%%%%%%
\section{Summary}
In this work we have improved a non local version of the Thirring model which provides
a tractable field-theoretical description of Luttinger liquids with spin. Indeed, in
the context of non local Thirring-like theories previous treatments of SF interactions
led to an involved non Abelian model (See for instance \cite{NLT1}). Specifically we
have built an action based on two fermion species which allows to take into account SF
interactions in an elegant and simpler way. Although our model was inspired by the one
considered in Ref.\cite{Zinn-Justin}, it includes interaction diagrams not contained
in that previous work (the so called $g_4$ diagrams in Solyom's terminology
\cite{Solyom}). Besides, the theory we present has general bilocal potentials
governing the interactions that do not flip spins. We parametrized this potentials in
terms of the functions $V^c$ and $V^s$ which become associated to charge-density and
spin-density dynamics respectively, once the spin-charge separation is made manifest
after suitable changes of variables (See equations (\ref{scSep1}) and (\ref{scSep2})).
On the other hand, we could not get closed formulae for arbitrary distance-dependent
SF potentials. Thus our analysis is only valid for local (contact) SF interactions.
However, we were able to keep the general distance dependence of the ordinary (no SF)
potentials up to the end of our computations. Under these conditions we got an
effective bosonic action whose charge-density piece coincides with previously obtained
descriptions of the forward-scattering spinless problem (\cite{NLT1}. Concerning the
more interesting spin-charge sector, we obtained, within the framework of the SCHA, a
general equation which allows to have the gap in terms of arbitrary electron-electron
potentials. Using this formula we could, for instance, estimate the effect of a
long-range interaction on the gap. \vspace{0.5cm}

{\bf Acknowledgements}

This work was supported by the Consejo Nacional de Investigaciones Cient\'{\i}ficas y
T\'ecnicas (CONICET) and Universidad Nacional de La Plata (UNLP), Argentina. KL has
benefited from a fellowship of CONICET and the Third World Academy of Sciences (TWAS).

%%%%%%%%%%%%%%%%%%%%%%%%%%%%%%%%%%%%%%%%%%%%%%%%%%%%%%%%%%%%%%%%%%%%%%%%%%%%%%%%%%%%%%%%%%%%%%%%%%%%%%%%%%%%%%%%%%%%%%
%%%%%%%%%%%%%%%%%%%%%%%%%%%%%%%%%%%%%%%%%%%%%%%%%%%%%%%%%%%%%%%%%%%%%%%%%%%%%%%%%%%%%%%%%%%%%%%%%%%%%%%%%%%%%%%%%%%%%%
\section{Appendix: details of the SCHA method}

We shall give an sketch of the SCHA method. One usually starts from a partition
function

\begin{equation}
\Z\true=\int\D\mu e^{-S\true}
\end{equation}

\noindent where $\D\mu$ is a generic integration measure. An elementary manipulation
leads to

\begin{equation}\label{eq:Z}
\Z\true=\frac{\int\D\mu e^{-(S\true-S\trial)}\,e^{-S\trial}}{\int\D\mu
e^{-S\trial}}\int\D\mu e^{-S\trial}=\Z\trial\bra e^{-(S\true-S\trial)}
\ket\trial.
\end{equation}

\noindent for any trial action $S\trial$. Now, by means of the property

\begin{equation}
\bra e^{-f} \ket\geq e^{-\bra f \ket},
\end{equation}

\noindent for $f$ real, and taking natural logarithm in equation (\ref{eq:Z}), we
obtain Feynman's inequality \cite{Feynman}

\begin{equation}\label{eq:Feynman}
\ln\Z\true\geq \ln\Z\trial - \bra S\true-S\trial\ \ket\trial
\end{equation}

In our case, the true action is the corresponding to the spin-charge dynamics, $S_s$.
However, before proceeding to the actual computation it is convenient to consider
still another change of variables which diagonalizes the cuadratic part of $S_s$. This
change is given by

\begin{equation}\label{eq:changeVariables}
\theta=\frac{ig}{\sqrt{\pi}}\,\zeta+\frac{A-\frac{C^2}{4B}}
{A-\frac{C^2}{4B}-\frac{g^2p^2}{2\pi}}\,\xi
\end{equation}

\begin{equation}
\eta=\frac{C}{2B}\,\zeta - \frac{i\sqrt{\pi}gCp^2}
{4B\left(A-\frac{C^2}{4B}-\frac{g^2p^2}{2\pi}\right)}\,\xi+\varphi
\end{equation}

\begin{equation}
\phi=-\zeta+\frac{igp^2}{\sqrt{4\pi}\left(A-\frac{C^2}{4B}-\frac{g^2p^2}{2\pi}\right)}\,\xi
\end{equation}
and the resulting action,  reads

\begin{multline}
S_s=\frac{1}{2}\int\difp\Bigg\{\zeta(p)\zeta(-p)\frac{p^4}{p_0^2 V^s_{(1)}+p_1^2 V^s_{(0)}}
+\varphi(p)\varphi(-p)\left(\frac{p_0^2}{V^s_{(0)}}+\frac{p_1^2}{V^s_{(1)}}\right)\\
+\xi(p)\xi(-p)\left[p^2+\frac{g^2}{\pi}\left(p_0^2 V^s_{(1)}+p_1^2 V^s_{(0)}\right)\right]\Bigg\}
-\frac{g_s\Lambda^2}{\pi^2}\int d^2x\,\cos(\sqrt{8\pi}\xi)
\end{multline}
where one sees that the $\zeta$ and $\varphi$ fields become completely decoupled from
$\xi$. Then it becomes apparent that the $\xi$-dependent piece is the relevant one to
consider in the present approximation. In other words, the relevant true and trial
actions are given by

\begin{equation}
S\true=\int\difp\xi(p)\frac{F(p)}{2}\xi(-p) - \frac{g_s\Lambda^2}{\pi^2}\int
d^2x\,\cos(\sqrt{8\pi}\xi)
\end{equation}

\begin{equation}
S\trial=\int\difp\left[\xi(p)\frac{F(p)}{2}\xi(-p) +
\frac{8\pi\Omega^2}{2}\xi(p)\xi(-p)\right].
\end{equation}

\noindent where

\begin{equation}
F(p)=p^2+\frac{g^2}{\pi}\left(p_0^2 V^s_{(1)}+p_1^2 V^s_{(0)}\right).
\end{equation}

The parameter $\Omega$ can be determined by maximizing the right hand side of equation
(\ref{eq:Feynman}). In order to achieve this goal we first write

\begin{align}
\ln Z\trial=&\ln\int\D\xi\exp{\left[-\frac{1}{2}\int d^2x\,\xi(x)
(\hat{A}\xi)(x)\right]} = [\ln(\det\hat{A})^{-1/2}]+\text{const}
\\=&-\frac{1}{2}\tr\ln\hat{A} + \text{const}.
\end{align}
where the operator $\hat{A}$ is defined, in Fourier space, by

\begin{equation}
(\hat{A}\xi)(p)=[F(p)+8\pi\Omega^2]\xi(p).
\end{equation}
It is then easy to get

\begin{equation}
\tr\ln\hat{A}=V\int\frac{d^2p}{(2\pi)^2} \ln[F(p)+8\pi\Omega^2]
\end{equation}
where $V$ is the volume (infinite) of the whole space $\int d^2x$. On the
other hand, it is straightforward to compute $\langle S\true-S\trial \rangle$, by following,
for instance, the steps explained in ref. \cite{Li-Naon}. The result is

\begin{multline}
-\bra S\true-S\trial \ket\trial = V\frac{g_s\Lambda^2}{\pi^2}
\exp\left[-4\pi\int\difp\frac{1}{F(p) + 8\pi\Omega^2}\right] \\+ V
4\pi\Omega^2\int\difp\frac{1}{F(p) + 8\pi\Omega^2}.
\end{multline}
Finally we can gather all the terms, and write them as

\begin{multline}\label{eq:elegant}
\ln\Z\trial - \bra S\true-S\trial \ket\trial\\=V\left[\frac{g_s\Lambda^2}{\pi^2}
\exp\left(-4\pi I_0(\Omega)\right) + \frac{8\pi\Omega^2}{2}I_0(\Omega) -
\frac{1}{2}I_1(\Omega)\right] + \text{const}.
\end{multline}

\noindent where we have defined the integrals

\begin{equation}
I_1(\Omega)=\int \frac{d^2p}{(2\pi)^2} \ln[F(p) + 8\pi\Omega^2]
\end{equation}

\begin{equation}
I_{-n}(\Omega)=\int\frac{d^2p}{(2\pi)^2}\frac{1}{\left[F(p) +
8\pi\Omega^2\right]^{n+1}}
\end{equation}

\noindent with the formal properties

\begin{equation}
\frac{dI_1(\Omega)}{d\Omega}=16\pi\Omega I_0
\end{equation}

\begin{equation}
\frac{dI_{-n}(\Omega)}{d\Omega}=-16\pi(n+1)\Omega I_{-n-1}.
\end{equation}

\noindent Now extrimizing expression (\ref{eq:elegant}) with respect to $\Omega$, and
assuming that $I_{-1}(\Omega)$ is not zero (a condition that holds for most realistic
potentials), we finally obtain the gap equation

\begin{equation}\label{eq:omegaeq}
\Omega^2-\frac{g_s\Lambda^2}{\pi^2} e^{-4\pi I_0(\Omega)}=0.
\end{equation}

\end{document}